\documentclass[prx,aps,amsmath,amssymb,showpacs,twocolumn]{revtex4-2}
\usepackage{graphicx,amsmath,amssymb,color}
\usepackage[utf8]{inputenc}
\usepackage{nicefrac}
\usepackage{multirow, makecell, ragged2e}
\usepackage[titletoc,title]{appendix}
\usepackage{physics}

\newcommand{\be}{\begin{equation}}
\newcommand{\ee}{\end{equation}}

\newcommand{\ba}{\begin{eqnarray}}
\newcommand{\ea}{\end{eqnarray}}

\renewcommand{\vec}[1]{\mbox{\boldmath$#1$}}

\def\beq{\begin{eqnarray}}
\def\eeq{\end{eqnarray}}
\newcommand{\sh}{\mathcal{S}}

\usepackage[colorlinks,bookmarks=true,citecolor=blue,linkcolor=red,urlcolor=blue]{hyperref}

\newcommand{\elliptic}[5][\scriptstyle]{\vartheta\left[\begin{array}{c}{{#1 #2}}\\{#1 #3}\end{array}\right]\left(#4\middle|#5\right)}

\begin{document}

\title{Fractional quantum Hall effect with unconventional pairing in monolayer graphene}
\author{Anirban Sharma$^1$, Songyang Pu$^{1,2}$, Ajit C. Balram$^{3,4}$, and J. K. Jain$^1$}
\affiliation{$^1$Department of Physics, 104 Davey Lab, Pennsylvania State University, University Park, Pennsylvania 16802, USA}
\affiliation{$^2$School of Physics and Astronomy, University of Leeds, Leeds LS2 9JT, United Kingdom}
\affiliation{$^{3}$Institute of Mathematical Sciences, CIT Campus, Chennai, 600113, India}
\affiliation{$^{4}$Homi Bhabha National Institute, Training School Complex, Anushaktinagar, Mumbai, 400094, India}

\date{\today}

\begin{abstract}
Motivated by the observation of even denominator fractional quantum Hall effect in the $n=3$ Landau level of monolayer graphene [Y. Kim {\em et al.}, Nature Physics {\bf 15}, 154 (2019)], we consider a Bardeen-Cooper-Schrieffer variational state for composite fermions and find that the composite-fermion Fermi sea in this Landau level is unstable to an $f$-wave pairing. Analogous calculation suggests the possibility of a $p$-wave pairing of composite fermions at half filling in the $n=2$ graphene Landau level, whereas no pairing instability is found at half filling in the $n=0$ and $1$ graphene Landau levels. The relevance of these results to experiments is discussed.
\end{abstract}
\maketitle

The $\nu=5/2$ fractional quantum Hall effect (FQHE) at half-filled second Landau level (LL) in semiconductor quantum wells \cite{Willett87} has been modeled through a Moore-Read (MR) Pfaffian wave function, which represents a $p$-wave paired state of the spin-polarized composite fermions \cite{Moore91, Read00}, where the composite fermion (CF) is a topological particle composed of an electron and an even number of quantized vortices \cite{Jain89, Jain07}. This raises the question of whether CF pairs with other symmetry can also be realized. 

Which FQHE state occurs depends on the Haldane pseudopotentials $V_{m}$~\cite{Haldane83} ($V_{m}$ is the energy of two electrons in a state with relative angular momentum $m$), which, in turn, are determined by both the interaction and the LL in which the electrons reside.  Graphene provides a platform for the realization of many old as well as new FQHE states. Unexpectedly, an FQHE state has been observed at half-filling in the $n=3$ LL of monolayer graphene \cite{Kim19}.  Ref.~\cite{Kim19} considered many candidate FQHE states and concluded that while none matches the Coulomb ground state, the 221 parton state~\cite{Jain89b} is the most promising because it can be stabilized when the $V_1$ and $V_3$ pseudopotentials are varied slightly away from their pure Coulomb values (which may in principle happen due to screening by metallic gates or LL mixing).  A realization of this state would be of interest because it represents an $f$-wave pairing of composite fermions~\cite{Balram18, Faugno19} and supports Ising type non-Abelian quasiparticles~\cite{Wen91, Bandyopadhyay18}. It is also the exact ground state~\cite{Wu17,Bandyopadhyay18} for the short range Trugman-Kivelson model interaction~\cite{Trugman85}. The 221 and the related 22111 states have been shown theoretically to be promising candidates also for $1/2$ FQHE in multilayer graphene~\cite{Wu17} and $1/4$ FQHE~\cite{Faugno19} observed in wide quantum wells~\cite{Shabani09a, Shabani09b, Shabani13}.

We investigate in this work the possibility of CF pairing in monolayer graphene directly from the Bardeen-Cooper-Schrieffer (BCS) perspective. Such an approach has previously been used in the contexts of $p$-wave CF pairing in the 5/2 state~\cite{Moller08,Sharma21} and $s$ and $p$-wave CF pairing in bilayer systems~\cite{Wagner21}. We consider more general pairings to address even-denominator FQHE in graphene. We employ a BCS wave function of composite fermions in the torus (periodic) geometry, which is convenient for momentum space pairing~\cite{Sharma21}.  This wave function has two variational parameters, analogous to the gap function and the Debye cutoff of the standard BCS theory.  An advantage of this method is that it enables a study of the competition between different kinds of pairing instabilities. Specifically, we can choose the gap function as $\Delta^{(l)}_{\vec{k}}\sim e^{-il\theta}$, where $\theta$ is the angular coordinate of the wave vector $\vec{k}$, and the relative angular momentum $l$ must be an odd integer for fully spin-polarized fermions. The choice $l=1$ corresponds to $p$-wave pairing and $l=3$ to $f$-wave (in our convention of magnetic field $B$ pointing in the $-\vec{z}$ direction). Another advantage of this method is that it allows minimization of energy by adjusting parameters and thus may capture physics missed in studies that use a single, fixed wave function. Finally, the various paired states are explicitly seen to arise through an instability of the CF Fermi sea (CFFS), which is a special case of the CF-BCS wave function. Ref.~\cite{Sharma21} demonstrated that this approach is capable of capturing the $p$-wave pairing instability at $\nu=5/2$ in semiconductor systems.

We find that the CFFS is unstable to $f$-wave pairing at half filling in the $n=3$ graphene LL. Notably, this instability is seen without any modification to the Coulomb interaction. No pairing instability occurs in the $n=0$ or $n=1$ graphene LL, but our work suggests the possibility of $p$-wave pairing in the $n=2$ graphene LL.

Our starting point is the BCS wave function for composite fermions on a torus. We consider a torus defined by a parallelogram with sides $L$ and $L\tau$, where the complex number $\tau=\tau_1+i\tau_2$ specifies the aspect ratio (or the modular parameter) of the torus \cite{Gunning62}.  The allowed values of wave vectors are
$\vec{k} = \left[n_1+{\phi_1\over 2\pi} \right]\vec{b_1} + \left[n_2+{\phi_2\over 2\pi} \right] \vec{b_2}$, with
$\vec{b}_1=\left({2\pi\over L},-{2\pi\tau_1\over L\tau_2}\right),\;\;
\vec{b}_2=\left(0,{2\pi\over L\tau_2}\right)$, where the angles $\phi_j$ represent phase twists in quasiperiodic boundary conditions. We take $\phi_1=\phi_2=\pi$ in what follows, to ensure that $\vec{k}=0$ is not an allowed value, and for each $\vec{k}$, $-\vec{k}$ is also allowed.  We define $z_j=x_j+iy_j$, where $\vec{r}_j\equiv(x_j,y_j)$ are the coordinates of the $j$th electron.

The BCS wave function for fully spin-polarized electrons is written as
$\ket{\Psi_{\rm BCS}} = \prod'_{\vec{k}}(u_{\vec{k}} + v_{\vec{k}}c_{\vec{k}}^{\dagger}c_{-\vec{k}}^{\dagger})\ket{0}$, where $\ket{0}$ is the null state, $c_{\vec{k}}^{\dagger}$ creates an electron at the wave vector $\vec{k}$, and each $\vec{k},-\vec{k}$ is counted only once in the product, and $|v_{\vec{k}}|^{2}$ ($|u_{\vec{k}}|^{2}$) is the probability of this state to be occupied (empty).  The real space form of the BCS wave function for a fixed number of electrons is given by~\cite{DeGennes99}
\begin{equation}
\Psi_{\rm BCS}(\vec{r}_1,...\vec{r}_N) = {\rm Pf}\left[g^{(l)}(\vec{r}_i-\vec{r}_j)\right],
\end{equation}
where Pf refers to Pfaffian, and the antisymmetric matrix element $g^{(l)}(\vec{r}_i-\vec{r}_j)$ can be expanded as
\begin{equation}
g^{(l)}(\vec{r}_i-\vec{r}_j) = \sum_{\vec{k}}g^{(l)}_{\vec{k}} e^{i{\vec{k}}\cdot(\vec{r}_i-\vec{r}_j)}
\label{eq:glr}
\end{equation}
with
\be
\label{gk}
g^{(l)}_{\vec{k}}\equiv{v_{\vec{k}}}/{u_{\vec{k}}}= \left[{\epsilon_{\vec{k}} - \sqrt{\epsilon_{\vec{k}}^2 +|\Delta^{(l)}_{\vec{k}}|^2}}\right]/{\Delta^{(l)*}_{\vec{k}}}=-g^{(l)}_{-\vec{k}}.
\ee
Here $\epsilon_{\vec{k}}=\hbar^2 (|\vec{k}|^2-k_F^2)/2m_{e}$ (we determine the magnitude of $k_F$ using the relation:
$\pi |k_F|^2 = N |\vec{b_1} \cross \vec{b_2}|$), and the gap function for the $l$ pairing channel has the form $\Delta^{(l)}_{\vec{k}}=\Delta |\vec{k}|e^{-il\theta}$, where $\theta$ is the angular coordinate of $\vec{k}$, with $l=1$ and $l=3$ corresponding to  $p$-wave and $f$-wave pairing (This form corresponds to the real space pair wave function of the form $e^{il\theta}/|z_i-z_j|$ for large $|z_i-z_j|$.). We note that we can alternatively choose $\Delta^{(l)}_{\vec{k}}=\Delta |\vec{k}|^le^{-il\theta}$. The two choices are equivalent in the limit where only wave vectors on the Fermi surface are relevant to pairing, in which case $|\vec{k}|$ can be replaced by $k_F$ (Our explicit calculations shown in the Supplemental Material (SM)~\cite{SM-Sharma22} demonstrate that the conclusions are not affected by this detail.).

\begin{figure}[t]
\vspace*{-3mm}
\includegraphics[width=0.3\linewidth]{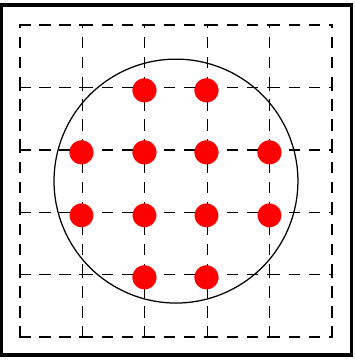}
\includegraphics[width=0.3\linewidth]{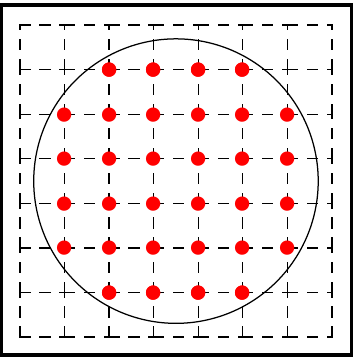}

\caption{\label{Fermi_sea}  Fermi seas for $N=12$ and 32 composite fermions.
}
\end{figure}

The BCS wave function for CFs at $\nu=1/2$ can now be constructed in the standard manner by vortex attachment \cite{Jain89, Jain07}. In the disk geometry, one would write $\Psi^{\rm CF-BCS}\sim P_{\rm LLL}{\rm Pf}[g^{(l)}(\vec{r}_i-\vec{r}_j)] \prod_{j<k}(z_j-z_k)^2$, where $P_{\rm LLL}$ refers to lowest-LL (LLL) projection. One would then attempt to implement the Jain Kamilla (JK) projection into the LLL \cite{Jain97,Jain97b} by writing the Jastrow factor as $\prod_{j<k}(z_j-z_k)^2=\prod_i J_i$, where $J_i=\prod_{k\neq i}(z_i-z_k)$;
incorporating it into the Pfaffian as  $\Psi^{\rm CF-BCS}\sim P_{\rm LLL}{\rm Pf}[g^{(l)}(\vec{r}_i-\vec{r}_j)J_iJ_j]$~\cite{Mishmash18}; and then projecting each matrix element separately into the LLL. In the torus geometry, we write
\be
\Psi _{\frac{1}{2}}^{\rm CF-BCS} = P_{\rm LLL} {\rm Pf}\left(\sum _{\vec{k}} g^{(l)}_{\vec{k}} e^{i\vec{k}_n\cdot\left(\vec{r}_i-\vec{r}_j\right)}\right) \Psi^{\rm L}_{1/2},
\ee
where $\Psi^{\rm L}_{1/2}$ is the $\nu=1/2$ Laughlin wave function~\cite{Laughlin83} in the torus geometry~\cite{Haldane85,Haldane85b,Pu20b}, while also replacing the mass of electron $m_{e}$ in Eq.~\eqref{gk} by the CF effective mass of $m^*$.  An implementation of the standard JK projection \cite{Jain97, Jain97b} in the torus geometry yields unphysical wave functions that do not satisfy the stipulated periodic boundary conditions (PBC). However, a modified JK projection accomplishes the task~\cite{Pu17, Pu18, Sharma21}. The resulting LLL wave function has the form (see Ref.~\cite{Sharma21} and SM~\cite{SM-Sharma22} for details)
\begin{widetext}
\begin{equation}
\label{BCS_paired}
    \Psi_{\frac{1}{2}}^{\rm CF-BCS} = e^{\sum_i \frac{z_i^2 - |z_i|^2}{4\ell^2}}\Bigg\{\vartheta
\begin{bmatrix}
{\phi_1\over 4\pi}
 + {N_{\phi}-2 \over 4}\\
-{\phi_2\over 2\pi } + {N-1}
\end{bmatrix}
\Bigg({2Z \over L} \Bigg |2 \tau \Bigg) \Bigg \} {\rm Pf}(M_{ij})
\end{equation}
\beq
\label{LLL pfa}
    M_{ij} = \sum_{\vec{k}}g_{\vec{k}}e^{-\frac{\ell^2}{2}k(k+2\Bar{k})}e^{\frac{i}{2}(z_i-z_j)(k+\Bar{k})}\Bigg(\vartheta  \begin{bmatrix}
{1 \over 2} \\ {1 \over 2 }
\end{bmatrix}\Bigg(\frac{z_i +ik\ell^2 - (z_j-ik\ell^2)}{L}|\tau \Bigg)\Bigg)^2 \nonumber \\ \Bigg \{ \prod_{\substack{r \\r \neq i,j}}   \vartheta  \begin{bmatrix}
{1 \over 2} \\ {1 \over 2 }
\end{bmatrix}\Bigg(\frac{z_i + i2k\ell^2- z_r}{L}|\tau \Bigg) \prod_{\substack{m \\m \neq i,j}}  \vartheta  \begin{bmatrix}
{1 \over 2} \\ {1 \over 2 }
\end{bmatrix}\Bigg(\frac{z_j - i2k\ell^2- z_m}{L}|\tau \Bigg) \Bigg \}.
\eeq
\end{widetext}
Here $Z=\sum_{i=1}^Nz_i$ is the center-of-mass (COM) coordinate, $k=k_x+ik_y$, $\ell=\sqrt{\hbar c/eB}$ is the magnetic length, $N$ is the number of particles, and $N_\phi=2N$ is the number of flux quanta through the torus.
The Jacobi theta function with rational characteristics is defined as \cite{Mumford07}
$
\vartheta  \begin{bmatrix}
a \\ b
\end{bmatrix}(z|\tau) =\sum_{n = -\infty}^{\infty} e^{i\pi(n+a)^2 \tau} e^{i2 \pi(n+a)(z+b)}$.
The above BCS wave function satisfies proper quasiperiodic boundary conditions on the torus. In terms of a dimensionless ``gap parameter"  $\tilde{\Delta}=|\Delta^{(l)} _{k_F}| /( \hbar^2|k_F|^2/2m^*)$, we have $g^{(l)}_{\vec{k}} ={\abs{\vec{k}}^2 - \abs{k_F}^2-\sqrt{(\abs{\vec{k}}^2 - \abs{k_F}^2)^2+\tilde{\Delta}^2|k_F|^2\abs{\vec{k}}^2}\over \tilde{\Delta}\abs{\vec{k}}k_Fe^{il\theta}}$.  We introduce an additional variational parameter, namely a momentum cutoff $k_{\rm cutoff}$, analogous to the Debye cutoff of the BCS theory, by setting $g^{(l)}_{\vec{k}}=0$ for $|\vec{k}|>k_{\rm cutoff}$. For $k_{\rm cutoff}=k_F$ the CF-BCS wave function reduces to the CFFS wave function~\cite{Rezayi00,Pu18,Pu20b}.
The 221 state lies in the sector with Haldane pseudomomenta $(K_x,K_y)=(N/2,N/2)$ or $(K_x,K_y)=(0,N/2)$ or $(K_x,K_y)=(N/2,0)$~\cite{Sharma22unpub}; in what follows, we will choose our CF-BCS state in the sector $(K_x,K_y)=(N/2,N/2)$.

In the absence of LL mixing, the electron-electron interaction in the $n=0$ LL of monolayer graphene is identical to that in the LLL of GaAs quantum well with zero width. One, therefore, expects that the physics in the $n=0$ LL of monolayer graphene is identical to that at half-filled LLL in GaAs quantum well (of zero width), including the state at half-filling, which is well known to be a CFFS~\cite{Halperin93, Halperin20b, Shayegan20}. The interaction pseudopotentials in the $n\neq 0$ LLs of monolayer graphene are different from those in the corresponding LLs of semiconductor quantum wells. We numerically investigate the candidate states at half filling in the $|n|=1$, $|n|=2$, and $|n|=3$ LLs of graphene.

 \begin{figure*}[t]
\raisebox{0.02in}{\includegraphics[width=2.9in]{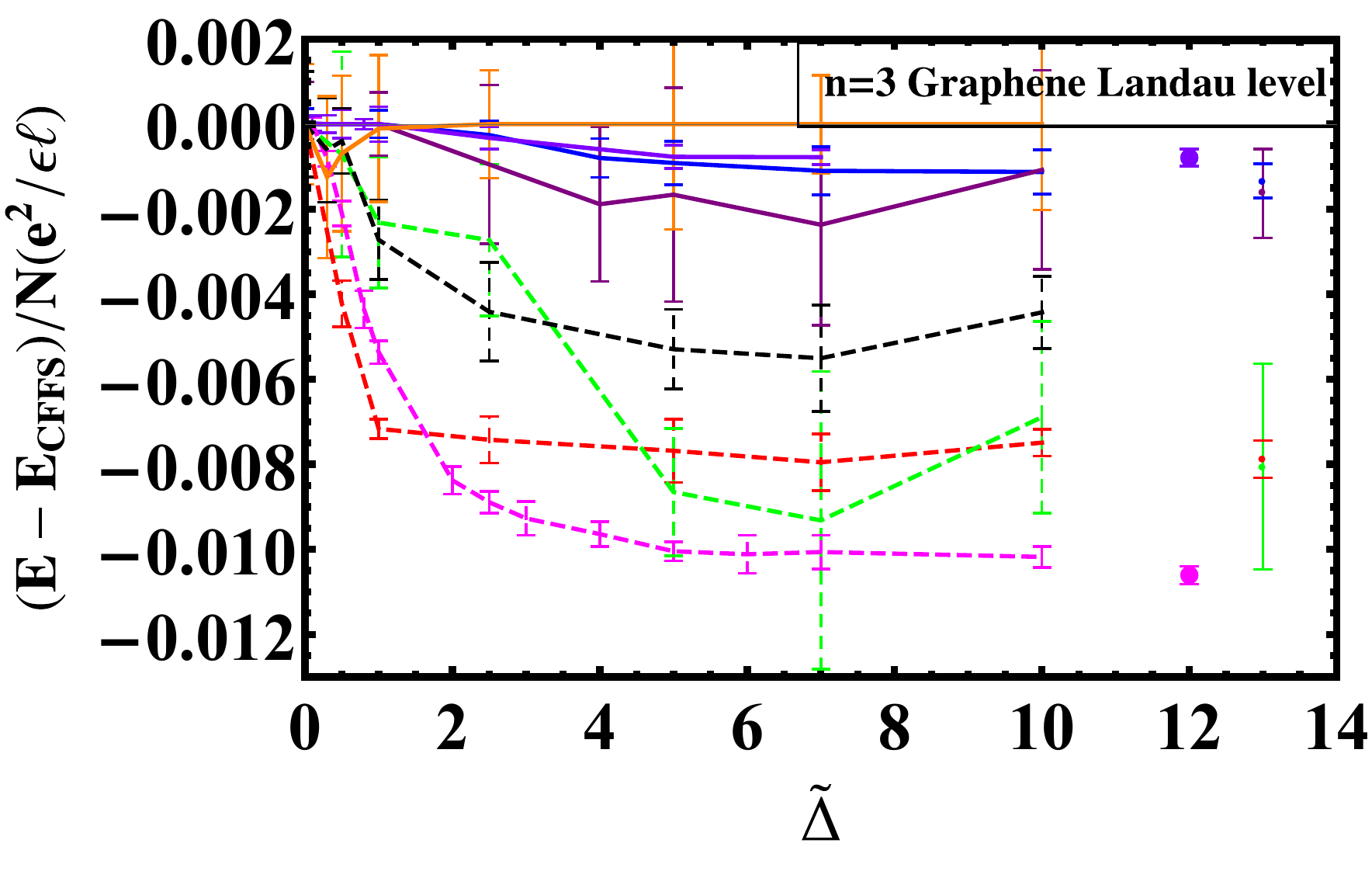}}
\raisebox{-0.03in}{\includegraphics[width=2.9in]{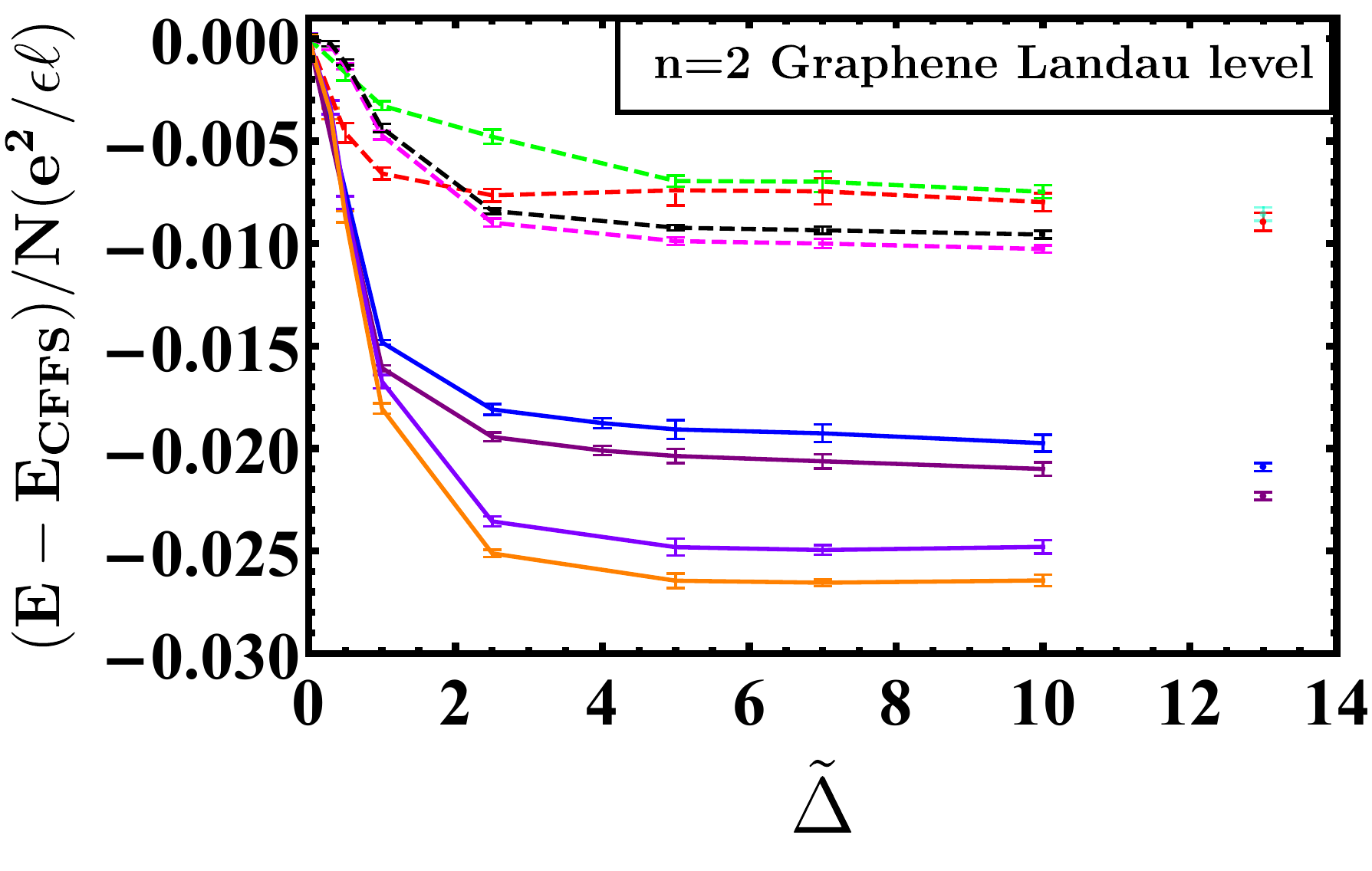}}
\raisebox{0.6in}{\includegraphics[width=1.0in]{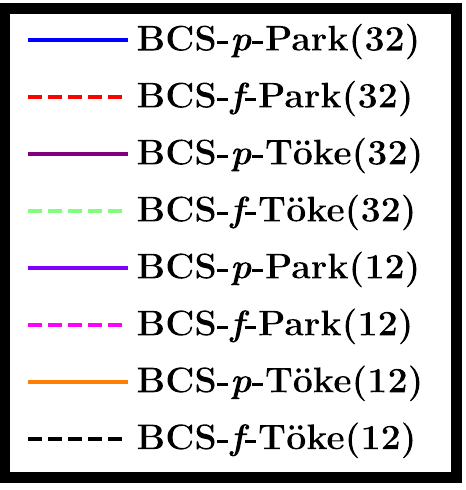}}
\vspace*{-3mm}
\caption{\label{Graphene_energy}  Energy per particle of the BCS-$p$ and BCS-$f$ wave states for the ``Park" and ``T\H oke" forms for the interaction (Refs.~\cite{Park98b,Toke05})  for the $n=3$ graphene LL (left panel) and $n=2$ graphene LL (middle panel) as a function of the gap parameter $\tilde{\Delta}$ for systems of 12 and 32 particles. The energies are quoted in units of $e^2/\epsilon \ell$, and measured relative to the energy of the CFFS. The energies for the $\tilde{\Delta}\rightarrow \infty$ limit are marked as isolated points at the right side of each plot. The legends are listed in the right panel, with the number of particles given in parentheses.
}
\end{figure*}

The inter-electron interaction in any given LL is completely specified by its Haldane pseudopotentials \cite{Haldane83}. The problem of electrons in the $n$th LL can thus be mapped into the problem of electrons in the $n=0$ LL with an effective interaction that has the same Haldane pseudopotentials as the Coulomb interaction in the $n$th LL. We consider two approximate real-space effective interactions\cite{Park98b,Toke05}: $V_{\rm Toke}=r^{-1}+\sum_{i=0}^6 c_i r^i e^{-r}$ and $V_{\rm Park}(r) = r^{-1} + a_1  e^{-\alpha_1 r^2} +a_2 r^2 e^{-\alpha_2 r^2}$.  For the former, we obtain the coefficients by matching the first seven odd pseudopotentials of the effective interaction ($V_{2m-1}$,$m=1,2,\cdots,7$) in LLL with the pseudopotentials of the Coulomb interaction in the $n$th graphene LL~\cite{Nomura06, Toke06, Toke07, Kim19}; for the latter, we match the first four odd pseudopotentials (in the $n=2$ LL, we need to make an additional approximation, discussed in the SM~\cite{SM-Sharma22}).  Both lead to the same conclusions, consistent with the expectation that the nature of the state is dictated by the first few odd pseudopotentials (even pseudopotentials are not germane for fully spin-polarized electrons). The validity of these effective interactions is further supported by the fact that the energy expectation values of the CF-BCS states for the Coulomb and the effective interactions are very nearly the same~\cite{SM-Sharma22}.
For the torus geometry, this interaction is replaced by an appropriate periodic interaction (see SM~\cite{SM-Sharma22} for details).
In the following, we assume spin-polarized electrons, disregard LL mixing, and quote all energies in units of $e^2/\epsilon \ell$.

We have calculated the energies of the CF-BCS wave function using the lattice Monte Carlo method \cite{Wang19}, which allows us to go to fairly large systems. We have considered systems with 12 and 32 particles because the Fermi seas for these systems are close to being circular (Fig.~\ref{Fermi_sea}).  We minimize the Coulomb energy in $n=2$ and $n=3$ graphene LLs with respect to the two variational parameters. Fig.~\ref{Graphene_energy} shows the minimum energy as a function of the gap parameter $\tilde{\Delta}$ where each point is obtained by minimizing the energy with respect to $k_{\rm cutoff}$. We note that because the CFFS is a special case of the BCS-$p$ and the BCS-$f$ states (with $k_{\rm cutoff}=k_F$), the minimum energy of the BCS-$p$ or the BCS-$f$ state is guaranteed to be less than or equal to that of the energy of the CFFS. Energy less than that of the CFFS implies a pairing instability of the CFFS.

As shown in Fig.~\ref{Graphene_energy}, the lowest energy state in $n=3$ graphene LL is obtained for the BCS-$f$ state.
Interestingly, the Coulomb energy is insensitive to the variation of the gap parameter $\tilde{\Delta}$ for larger values. In fact, the optimal state is well approximated by the limit $\tilde{\Delta}\rightarrow \infty$, where the CF-BCS state simplifies with $g_{\vec{k}}=-e^{-il\theta}$. 
The BCS-$p$ state may have slightly lower energy than the CFFS,  but has higher energy than the $f$-wave CF-BCS state.

\begin{figure}
\includegraphics[width=0.85\linewidth]{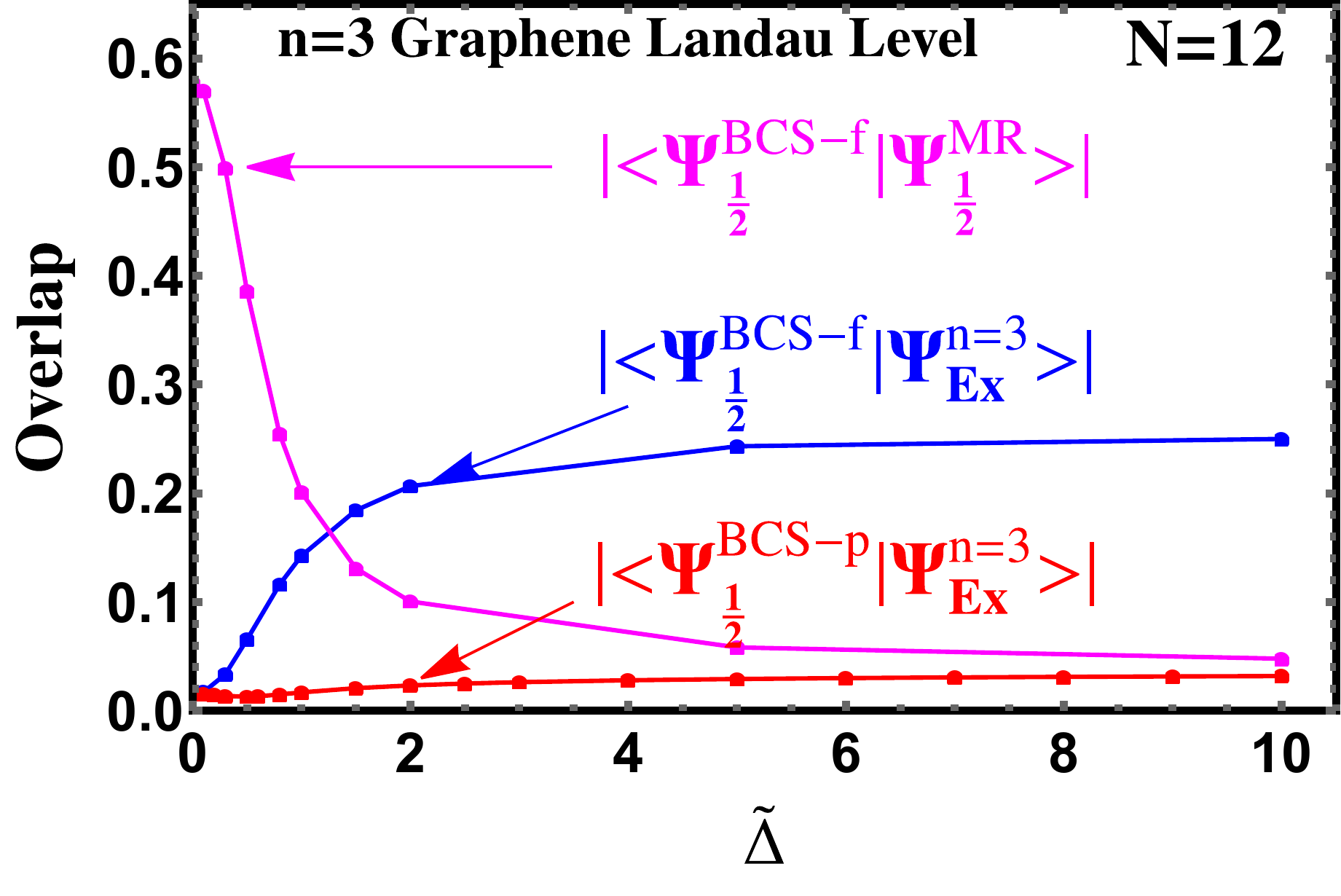}
\includegraphics[width=0.85\linewidth]{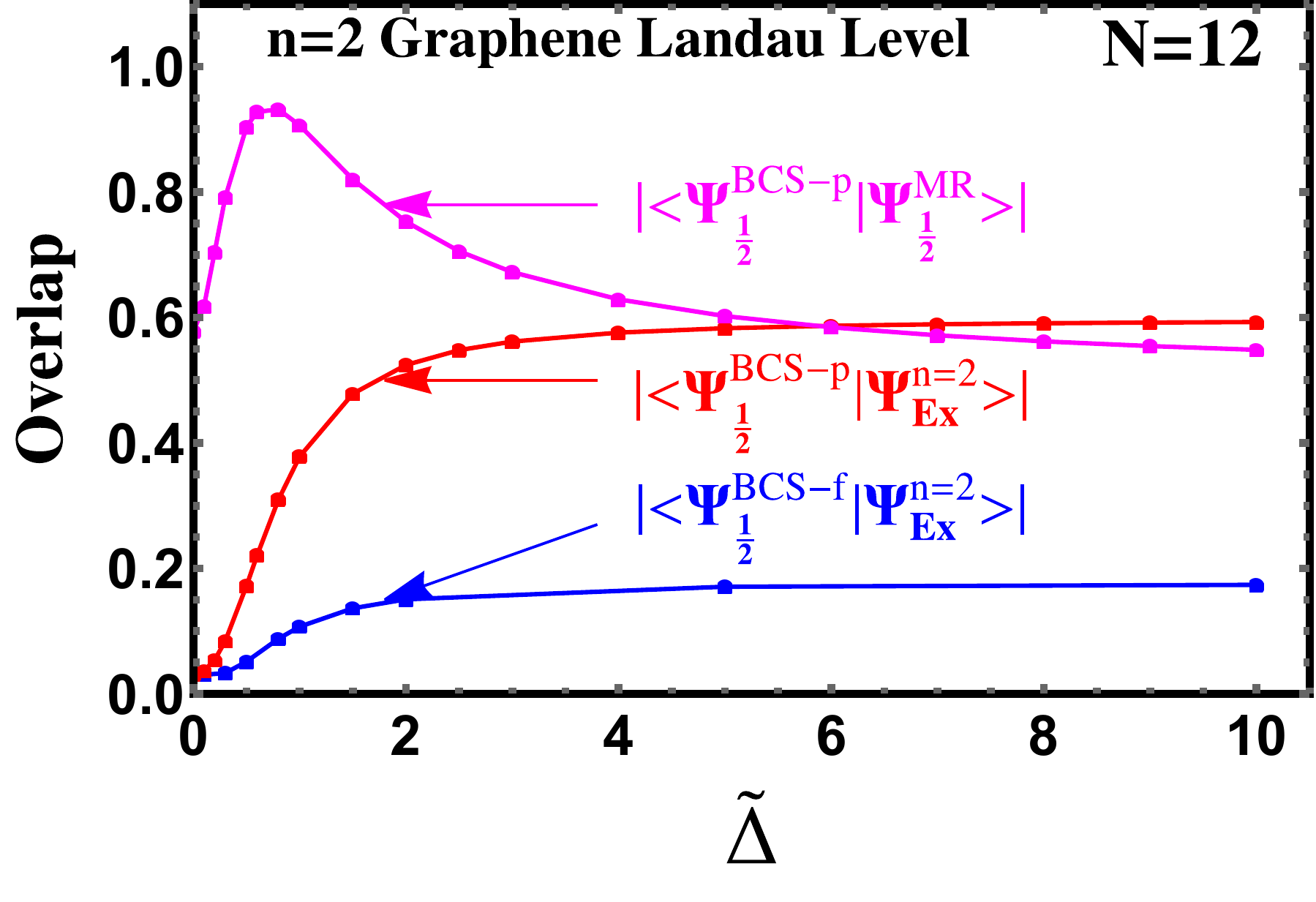}
\caption{\label{Graphene_2LL} Upper panel: Overlaps of the candidate states with the exact Coulomb ground state in the $n=3$ Landau level for 12 particles. The optimal overlap between the $f$-wave CF-BCS state and the exact ground state is approximately 0.25. The overlap of the $p$-wave paired state with the exact state is less than 0.025 for all values of $\tilde{\Delta}$. The overlap of the MR ($p$-wave) state with the exact Coulomb ground state in the $n=3$ LL is 0.01078. Lower panel: Overlaps of the candidate states with the exact Coulomb ground state in the $n=2$ Landau level of graphene for 12 particles. The optimal overlap between the $f$-wave CF-BCS state and the exact ground state is approximately 0.174. The optimal overlap of the $p$-wave paired state with the exact state is 0.5928. The overlap of the MR ($p$-wave) state with the exact Coulomb ground state in the n=2 LL of graphene is 0.19234.}
\end{figure}

Fig.~\ref{Graphene_2LL} shows the overlaps of the various candidate states with the exact ground state for the Coulomb interaction in graphene. For this purpose, we obtain the exact Fock-space representation of the CF-BCS state using the method in Ref.~\cite{Sreejith13}. The overlap of the exact Coulomb ground in graphene at the half-filled $n=3$ LL with BCS-$f$ state is approximately 0.25 in the parameter range where the energy is minimum. This overlap is not decisive, but still significant for an FQHE state in a high LL. [For the LLL, the wave functions of composite fermions at fractions $\nu=s/(2s\pm 1)$, $s$ integer, have overlaps of $\sim$0.99 with the Coulomb ground states for systems accessible to numerical diagonalization~\cite{Jain07, Yang19a, Balram21b}, but for the $n=1$ LL in GaAs the overlaps are generally much smaller; for example, the 7/3 and 5/2 Coulomb ground states have overlaps in the ranges 0.5-0.7 and 0.7-0.9, respectively, with the Laughlin and MR wave functions for numerically accessible particle numbers~\cite{Ambrumenil88, Morf98, Scarola02b, Balram13b, Balram20}.] The BCS-$f$ state is substantially better than other candidate states: the overlaps of the CFFS, BCS-$p$, and the MR-$p$ states with the exact Coulomb ground state at half filling in the $n=3$ graphene LL are, respectively,  0.01544, $\sim$0.025, and 0.01078 (we have used MR-$p$ wave function given in Refs.~\cite{Greiter92a,Chung07,Read96}).

In contrast to Ref.~\cite{Kim19}, the pure Coulomb interaction itself appears to produce CF pairing at half filling in the $n=3$ LL of graphene, thus providing important theoretical support to $f$-wave pairing. Given that our results for 12 and 32 particles are quite consistent, we speculate that for this problem, the torus geometry may better represent the thermodynamic behavior than the spherical geometry used in Ref.~\cite{Kim19}.  Exact diagonalization of the Coulomb interaction on torus with $\tau=i$ shows that the ground states for $N=8, 12, 14$ and $16$ particles lie in the sector with Haldane pseudomomenta $(K_x,K_y)=(N/2,N/2)$ or $(K_x,K_y)=(0,N/2)$ or $(K_x,K_y)=(N/2,0)$, which are the momentum sectors for the paired state~\cite{Sharma22unpub}. As shown in the SM, the BCS-$f$-wave state can be made stronger by modifying the interaction~\cite{SM-Sharma22}.

Fig.~\ref{Graphene_energy} also shows the results for $N=12$ and 32 particles at half filling in the $n=2$ graphene LL. We find that the lowest energy is obtained for the $p$-wave paired state. The overlap of the $12$ particle CF-BCS state with the exact Coulomb state is $\sim 0.6$ (Fig.~\ref{Graphene_2LL}). Intuitively, a $p$-wave pairing would not be entirely surprising here, as the $n=2$ graphene LL wave function is a combination of the $n=2$ and $n=1$ GaAs LL wave functions, the latter of which is believed to support $p$-wave pairing~\cite{Nomura06,Goerbig06,Morf98, Scarola02b, Balram20,Goerbig11}.

Our study does not decisively prove CF-pairing in $n=2$ and $n=3$ graphene LLs, as we have not ruled out all possible states such as the stripe phase. Nevertheless, we conclude that a paired state is at least competitive, and that {\it if} an FQHE state is observed at half filling in the $n=3$ ($n=2$) graphene LL, it is likely an $f$-wave (a $p$-wave) paired state. It is noted that $1/2$ FQHE has not yet been observed in the $n=2$ graphene LL~\cite{Diankov16, Chen19}; we have not considered the possibility of whether the paired state can be destabilized by LL mixing.

We find no pairing instability at half filling in the  $n=1$ graphene LL, i.e. our calculations show that the lowest energy is obtained when $k_{\rm cutoff}=k_{F}$ for arbitrary $\tilde{\Delta}$. This is in agreement with earlier variational and exact diagonalization studies~\cite{Toke07b,Wojs11a,Wojs11b,Balram15c, Balram21b}. The observation of many fractions along the sequences $\nu=s/(2s\pm 1)$ \cite{Amet15} is consistent with a CFFS at $\nu=1/2$.

The BCS-$f$ state is topologically distinct from the BCS-$p$ wave state. The thermal Hall conductance at temperature $T$, which is given by $\kappa = c \frac{\pi ^2 k_B^2}{3h} T$ where $c$ is the chiral central charge \cite{Kane97}, can in principle distinguish between them~\cite{Balram18, Faugno19, Rowell09}. The chiral central charge for different paired CF states is given by the relation $c=(1+l/2)$; in particular, for the $p$ and $f$ states considered here it is given by $c=3/2$ and $5/2$. The Hall viscosity $\eta^A$ \cite{Avron95} is given by \cite{Read09} $\eta^A=\hbar \rho \sh/4$, where $\rho$ is the 2D density and $\sh=N/\nu-N_\phi$ is the ``shift"  \cite{Wen92} in the spherical geometry.  For the $p$ and $f$ states at $\nu=1/2$ we have $\sh=3$ and $5$ respectively.

What is the mechanism of pairing? It is known empirically that a CFFS is obtained when the short distance repulsion between electrons is dominant, as is the case in the $n=0$ LL.  When the short distance repulsion is reduced, which is what happens in higher LLs, the effective interaction between composite fermions may become attractive, causing pairing. We do not have a simple way to predict which pairing is preferred without performing a detailed calculation. A Chern-Simons based analysis of gauge fluctuations, as in Ref.~\cite{Bonesteel96}, could provide further insight into this question.

In summary, we have minimized the energy of the CF-BCS wave function to determine the optimal pairing at half filling in graphene LLs. We find an absence of pairing instability in the $n=0$ and $n=1$ LLs but CF pairing appears possible in $n=2$ and $n=3$ LLs. Our primary conclusion is that if FQHE is observed in the $n=2$ ($n=3$) graphene LL, it likely represents $p$ ($f$) wave pairing of composite fermions. We hope that this study will motivate further experimental investigations of these states, which will be necessary for a definitive confirmation of their physical origin.

Acknowledgments:  A.S. and J.K.J thank the U.S. National Science Foundation for financial support under grant no. DMR-2037990. S.P. was supported (at Penn State) by the U.S. Department of Energy, Office of Basic Energy Sciences, under Grant no. DE-SC-0005042 and is supported (at Leeds) by the Leverhulme Trust Research Leadership Award RL-2019-015 and EPSRC grant EP/R020612/1. A. C. B. acknowledges the Science and Engineering Research Board (SERB) of the Department of Science and Technology (DST) for financial support through the Start-up Grant SRG/2020/000154. We have made use of Advanced CyberInfrastructure computational resources provided by The Institute for CyberScience at The Pennsylvania State University. We thank M. Wimmer for the open-source PFAPACK library~\cite{Wimmer12}, used for the numerical evaluation of the Pfaffian of matrices. We are grateful to the authors of the DiagHam package, which we used for some of our exact diagonalization calculations.

\renewcommand{\thesection}{S\arabic{section}}  
\renewcommand{\thetable}{S\arabic{table}}

\section*{Supplemental Material}

The supplementary material contains the following.  In Sec.~\ref{sec:review}, we provide a brief review, for completeness, of the basics of composite fermions on torus and also introduce certain known wave functions. In Sec.~\ref{wavefunction}, we review our CF-BCS wave function for spin-polarized composite fermions. Section \ref{sec:graphene} gives the interaction pseudopotentials for graphene LLs. Section \ref{appx-periodic-int} contains technical details of the energy calculation.  Certain consistency checks are described in Sec.~\ref{sec:check}. Finally, we provide a review of the lattice Monte Carlo approach along with certain details of our sampling procedure in Sec.~\ref{sec:Monte-rev}.

\section{Composite fermions on torus}
\label{sec:review}

This section contains a review of various relevant wave functions in the torus geometry. A torus can be mapped to a parallelogram with periodic boundary conditions \cite{Yoshioka83,Haldane85,Greiter16,Fremling14,Pu17,Pu18,Pu21,Pu20,Pu20b}.  The two edges of the parallelogram are denoted by $L_1=L$ and $L_2=L\tau$, where $\tau=\tau_1+i\tau_2$ is a complex parameter specifying the torus. $L_1$ is considered to be along the real axis. The magnetic field $\vec{B}=-B\hat{z}$ is perpendicular to the parallelogram. The complex coordinates of the particles are represented by $z=x+iy$. We work with the symmetric gauge given by $A=\frac{B}{2}(y,-x,0)$. The single particle wave functions on the torus satisfy the periodic boundary conditions in the two directions:
\begin{eqnarray}
t(L_1)\psi(z,\bar{z}) =  e^{i\phi_1 } \psi(z,\bar{z})\\ \nonumber
t(L_2)\psi(z,\bar{z}) =  e^{i\phi_2 } \psi(z,\bar{z})
\end{eqnarray}
where $t(L_i)$ is the magnetic translation operator in the $L_i$ direction.  
The magnetic translation operator $t(\xi)$ with $\xi = \xi _x + i \xi _y$ is 
defined by
\begin{equation}
t(\xi) =  e^{-\frac{i}{2\ell ^2}\hat{\vec{z}}.(\vec{\xi} \times \vec{r})} T(\xi)
\end{equation}
where $\ell=\sqrt{\hbar c/eB}$ is the magnetic length, and  
\begin{equation}
T(\xi) = e^{\xi \partial _z + \bar{\xi} \partial _{\bar{z}}}
\end{equation}
is the translation operator.
One can use 
\beq
t(L_1)e^{z^2-|z|^2 \over 4 \ell ^2} = e^{z^2-|z|^2 \over 4 \ell ^2} T(L_1)
\eeq
and
\beq
t(L_2)e^{z^2-|z|^2 \over 4 \ell ^2} = e^{z^2-|z|^2 \over 4 \ell ^2}e^{-i\pi N_{\phi}(2z/L+\tau)} T(L_2)
\eeq
to show that the many-particle wave functions satisfy the periodic boundary conditions
\ba
t_j(L_1)\Psi(\{z_i\},\{\bar{z}_i\})= e^{i\phi_1}\Psi(\{z_i\},\{\bar{z}_i\})  \\ \nonumber
t_j(L_2)\Psi(\{z_i\},\{\bar{z}_i\})= e^{i\phi_2}\Psi(\{z_i\},\{\bar{z}_i\})
\ea
where $t_j$ is the magnetic translation operator for the $j$th particle.

\subsection{Certain wave functions}

In this subsection, we list certain wave functions in the torus geometry that have been used in this work. These are expressed in terms of the Jacobi theta function with rational characteristics~\cite{Mumford07}, defined as
\be
\elliptic[\displaystyle]abz\tau=\sum_{n=-\infty}^{\infty}e^{i\pi \left(n+a\right)^2\tau}e^{i2\pi \left(n+a\right)\left(z+b\right)}.
\ee
It is useful to list here several periodic properties of Jacobi theta function that are used in demonstrating the quasiperiodicity of various wave functions:
\be
\elliptic[\displaystyle]ab{z+1}\tau=e^{i2\pi a}\elliptic[\displaystyle]abz\tau 
\ee
\be
\elliptic[\displaystyle]ab{z+\tau}\tau=e^{-i\pi [\tau+2(z+b)]}\elliptic[\displaystyle]abz\tau 
\ee
\be
\elliptic[\displaystyle]ab{z+w}{w\tau}=\elliptic[\displaystyle]a{b+w}z{w\tau} 
\ee
\be
\label{CP6 brunch}
\elliptic[\displaystyle]ab{z+\tau}{w\tau}=e^{-i{2\pi \over w}\left(z+b+{\tau\over 2}\right)}\elliptic[\displaystyle]{a+{1\over w}}bz{w\tau} 
\ee
where $w$ is a real number. 

{\it Laughlin wave function:}

The general Laughlin wave function~\cite{Laughlin83} can be written as \cite{Haldane85,Pu20b}
\begin{widetext}
\be
\label{Laughlin}
\Psi^{\rm L}_{1/m,k_{\rm CM}}[z_i,\bar{z}_i]=e^{\sum_i {z_i^2-\abs{z_i}^2 \over 4 \ell^2}}
\left[\elliptic{{\phi_1\over 2\pi m}+{k_{\rm CM}\over m}+{N-1\over2}}{-{\phi_2\over 2\pi}+{m(N-1)\over2}}{mZ \over L_1}{m\tau}\right] 
\prod_{i<j} \left[ \elliptic{{\frac12}}{{\frac12}}{z_{i}-z_{j}\over L_1}{\tau} \right]^m
\ee
where $k_{\rm CM}=0,1,\cdots m-1$ labels the eigenvalue under center-of-mass translation
\be
\prod_{i=1}^N t_i(L_1/N_\phi)\Psi^{\rm L}_{1/m,k_{\rm CM}}[z_i,\bar{z}_i]=e^{\i2\pi({\phi_1\over 2\pi m}+{k_{\rm CM}\over m}+{N-1\over2})}\Psi^{\rm L}_{1/m,k_{\rm CM}}[z_i,\bar{z}_i].
\ee
\end{widetext}

{\it CFFS wave function:}

The CFFS is given by 
\beq
\Psi^{\rm CFFS}_{k_{\rm CM}} =  P_{\rm LLL} {\rm Det}[e^{i\vec{k}_n\cdot\vec{r}_m}]\Psi_{1/2,k_{\rm CM}}^{\rm L} 
\label{CFFS1}
\eeq
where $P_{\rm LLL}$ is the lowest Landau level (LLL) projection operator; the details of LLL projection can be found in Refs.~\cite{Pu18,Pu20b}. The wave vectors $\vec{k}$ span the CFFS, and their allowed values constrained by the PBC are
\beq
\label{kn}
\vec{k}_n = \left[n_1+{\phi_1\over 2\pi} \right]\vec{b_1} + \left[n_2+{\phi_2\over 2\pi} \right] \vec{b_2}
\eeq
where
\be
\vec{b}_1=\left({2\pi\over L},-{2\pi\tau_1\over L\tau_2}\right),\;\;
\vec{b}_2=\left(0,{2\pi\over L\tau_2}\right).
\ee
The last part $\Psi_{1/2,k_{\rm CM}}^{\rm L} $ is the Laughlin wave function at $\nu=1/2$.

\section{CF-BCS wave function}
\label{wavefunction}

The LLL projected form of the CF-BCS wave function at $\nu=1/2$ can be expressed as
\begin{widetext}
\begin{equation}
 \Psi_{\frac{1}{2}}^{\rm CF-BCS} = e^{\sum_i \frac{z_i^2 - |z_i|^2}{4\ell^2}}\elliptic{{\phi_1\over 4\pi }+{N-1\over 2}}{-{\phi_2\over 2\pi}+(N-1)}{{2Z\over L_1}}{2\tau}{\rm Pf}\left[\sum_{\vec{k}}g^{(l)}_{\vec{k}}\hat{F}_k(z_i,z_j)\right]\prod_iJ_i,
\end{equation}
where $ J_i=\prod_{r \neq i} \vartheta 
\begin{bmatrix}
{1\over 2} \\ {1\over 2}
\end{bmatrix}
\Bigg( {z_i-z_r \over L}\Bigg | \tau \Bigg )  $
and 
\begin{equation}
\hat{F}_k(z_i,z_j)=e^{-\frac{k\ell^2}{2}(k+2\Bar{k})}e^{\frac{i}{2}(z_i - z_j)(k+\Bar{k})}e^{ik\ell^2 \partial _{z_i}} e^{-ik\ell^2 \partial _{z_j}} .
\end{equation}
The above form of the wave function is not amenable to calculations for large systems, because the LLL projection can be performed only for rather small systems. Following the standard Jain-Kamilla projection method~\cite{Jain97, Jain97b}, one can  
bring the Jastrow factor inside the Pfaffian matrix as
\begin{equation}
{\rm Pf}\left[\sum_{\vec{k}}g^{(l)}_{\vec{k}}\hat{F}_n(z_i,z_j)\right]\prod_iJ_i\rightarrow 
{\rm Pf}\left[\sum_{\vec{k}}g^{(l)}_{\vec{k}}\hat{F}_n(z_i,z_j)J_i J_j\right]
\label{method1}
\end{equation}
\end{widetext}
 This wave function, as it stands, does not satisfy the correct periodic boundary conditions. It was shown in Ref.~\cite{Sharma21} how the LLL projection can be modified to preserve the boundary conditions. The resulting wave function is given in the main text.

\section{Pseudopotentials for Graphene Landau levels}
\label{sec:graphene}

The Coulomb interaction can be described completely by its Haldane pseudopotentials $V_m$, which are the energies of two electrons in relative angular momenta $m$. The Haldane pseudopotentials for Coulomb interaction in the $n$th LL  in the planar geometry is given by: 
\begin{equation}
V^{(n)}_m = \int_{0}^{\infty} dq~F^{(n)}(q)e^{-q^2}L_{m}(q^2),
\end{equation}
where $F^{(n)}(q)$ is the form factor and $L_m(x)$ is the $m$-th order Laguerre polynomial. For the LLs of a parabolic system (as in GaAs quantum wells), the form factor is given by $F^{(n)}(q) =[L_n(q^2/2)]^2$. In graphene, the form factor is given by~ \cite{Nomura06,Toke06,Toke07,Kim19}:
\begin{equation}
F^{(n)}(q) = 
\begin{cases}
1 & n=0 \\
\frac{1}{4}\Bigg [L_{|n|-1}(\frac{q^2}{2})+L_{|n|}(\frac{q^2}{2})\Bigg]^2 & n \neq 0
\end{cases}
\end{equation}
The pseudopotentials for the Coulomb interaction in the $|n|=1$, $|n|=2$ and $|n|=3$ graphene Landau levels are given by\cite{Kim19,Balram15c}: 
\begin{widetext}
\begin{equation}
V^{(1)}_m=\Bigg (m^2 - \frac{15}{8} + \frac{153}{256} \Bigg ) \frac{\Gamma (m - \frac{3}{2})}{2 \Gamma (m+1)}
\end{equation}
\begin{equation}
V^{(2)}_m = \frac{(65536m^4 -499712m^3 + 1250048 m^2 - 1136032 m + 264705)\Gamma (m-\frac{7}{2})}{131072 ~\Gamma(m+1)}
\end{equation}
\begin{equation}
V^{(3)}_m = \frac{(64m(4m(16m(4m(32m(8m-139)+29817)-386923)+9915059)-27868989)+361610865)\Gamma (m-\frac{11}{2})}{8388608~ \Gamma(m+1)}
\end{equation}
\end{widetext}
For the $n=0$ graphene LL, the pseudopotentials are the same as in GaAs quantum wells. 
The problem of electrons in the $n$th LL is equivalent to that of electrons in the LLL interacting with an effective interaction that reproduces the desired pseudopotentials. We use effective interactions $V_{\rm Park}$ and $V_{\rm Toke}$ whose forms are given in the main text. To obtain the coefficients $c_i$ for $i=0,...,6$ in $V_{\rm Toke}$, we match the first seven odd pseudopotentials. The coefficients $a_1,a_2,\alpha_1,\alpha_2$ in $V_{\rm Park}$ are obtained by matching the first four odd pseudopotentials. For the $n=2$ LL in graphene, we obtain the parameters $a_1,a_2,\alpha_1,\alpha_2$  with $\delta V^{(2)}_5=0.0042$, where $\delta V^{(2)}_5$ is the change in the value of the fifth pesudopotential in $n=2$ LL; this ensures the existence of numerical solution to the system of equations. No such variation is needed for the other cases. The values of the coefficients of both the interactions are listed in Table \ref{coeff-table}.

\begin{center}
\begin{table}[t]
 \begin{tabular}{|c |c |c |c|} 
 \hline
Coefficient & $n=1$ & $n=2$ & $n=3$ \\
 \hline\hline
 $c_0$ & -6.631003 & -50.613975 &492.523594   \\
 \hline
 $c_1$ & 13.297839 & 76.637596 & -976.021237  \\
 \hline
$c_2$ &  -8.996789 & -42.305604 &  692.712510\\ 
\hline
 $c_3$ & 2.934041&  11.542866 &-235.341758 \\
  \hline
$c_4 $ & -0.498770 & -1.666281 & 41.446029\\ 
  \hline
$c_5$ & 0.042572 & 0.121853 &-3.644917 \\
 \hline
 $c_6$ & -0.001430 & -0.003521 &0.126428 \\
 \hline
 $a_1$ & 0.0107017 & 0.369253 & 11.8887 \\
 \hline
 $a_2$ & 0.109467 & -443.84 & -9.64883 \\
 \hline
 $\alpha _1$ & 0.038443 & 0.129292 & 0.247147 \\
 \hline
 
 $\alpha _2$ &  0.446909 & 6.23332 & 0.479972\\
 \hline
\end{tabular}

\caption{\label{coeff-table} Values of coefficients $c_i,a_i,\alpha_i$ of the effective interactions for $n=1,2$ and $3$ LL in graphene.}
\end{table}

\end{center}

\section{Interaction Energy}
\label{appx-periodic-int}

On torus, the interactions are periodic, i.e., satisfies
\beq
V(\vec{r}+mL_1+nL_2) = V(\vec{r})
\eeq
where $m$ and $n$ are integers. We use the periodic form of the interaction given by
\begin{equation}
\label{per-int}
V(\vec{r}) = \frac{1}{L^2\rm{Im} (\tau)}\sum_{\vec{q}} \tilde{V}(q)e^{i\vec{q}\cdot \vec{r}}
\end{equation}
\beq
\vec{q} = \left( {2\pi m \over L}, -{2\pi \tau _1 m \over L \tau _2} +{2\pi  n \over L \tau _2} \right)
\eeq
 where $\tilde{V}(\vec{q})$ is the Fourier transform of $V(\vec{r})$.
The Fourier transforms of the terms in the $V_{\rm Park}$ are:
\beq
\int\frac{1}{r} e^{-i\vec{q}\cdot\vec{r}} d^2 \vec{r}= \frac{2 \pi}{q}
\eeq
\beq
\int e^{-\alpha r^2} e^{-i\vec{q}\cdot\vec{r}}d^2\vec{r} =\left({\pi \over \alpha}\right) e^{-q^2\over 4 \alpha}
\eeq

\beq
\int \vec{r}^2 e^{-\alpha r^2} e^{-i\vec{q}\cdot\vec{r}}d^2\vec{r} = \left({\pi \over 2\alpha^2}\right) e^{-q^2\over 4 \alpha}\left(2-{q^2 \over 2 \alpha}\right)
\eeq
The effective interaction $V_{\rm Park}$ can thus be written as 
\beq
V_{\rm Park}(r)={1 \over L^2\rm{Im}(\tau)}\sum_{\vec{q}}\tilde{V}_{\rm Park}(q) e^{i\vec{q} \cdot \vec{r}}
\eeq
with 
\begin{equation}
\label{Park-q}
\tilde{V}_{\rm Park}(q) = {2\pi\over q} + a_1\left({\pi \over \alpha_1}\right) e^{-q^2\over 4 \alpha _1} + a_2\left({\pi \over 2 \alpha^2_2}\right) e^{-q^2\over 4 \alpha _2}\left(2-{q^2 \over 2 \alpha _2}\right).
\end{equation}
\vspace*{-3mm}
\begin{figure}[b]
		\includegraphics[width=\linewidth]{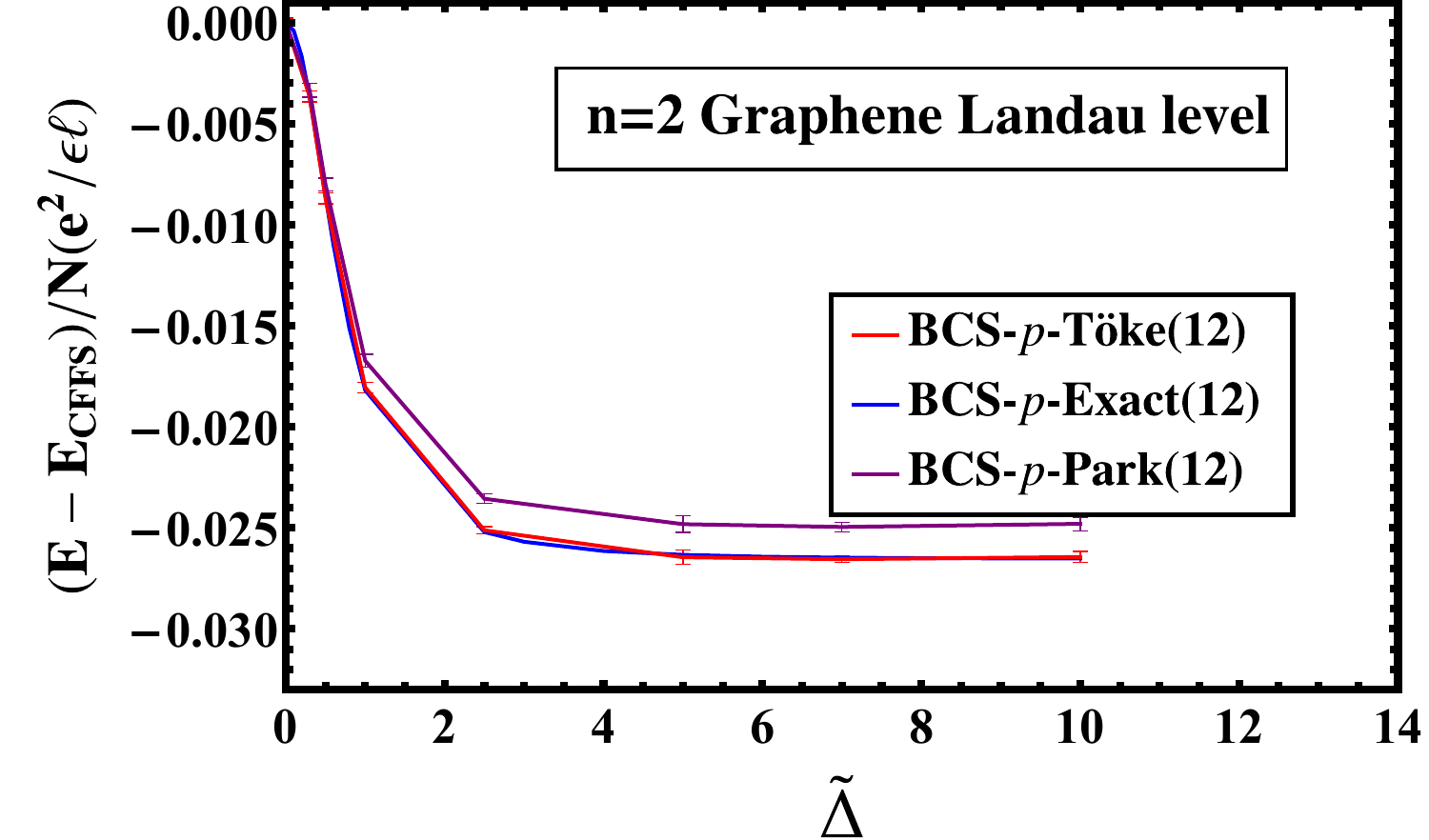}

\caption{ \label{exact-energy}Energy per particle for lowest energy $p$-wave paired CF-BCS state (at each $\tilde{\Delta}$) for $N=12$ particles in the $n=2$ LL of graphene using the Park and T\H oke interactions, as well as the exact Coulomb interaction. All energies are measured relative to the energy of the CFFS. 
}
\end{figure}
\vspace*{-3mm}

The Fourier transforms of the terms in $V_{\rm Toke}$ are given by:
\begin{gather}
\int e^{-r} e^{-i\vec{q}\cdot\vec{r}} d^2 \vec{r}= \frac{2 \pi}{(1+q^2)^{\frac{3}{2}}}\nonumber \\ 
\int re^{-r} e^{-i\vec{q}\cdot\vec{r}} d^2 \vec{r}= \frac{2 \pi(2-q^2)}{(1+q^2)^{\frac{5}{2}}} \nonumber \\ 
\int r^2e^{-r} e^{-i\vec{q}\cdot\vec{r}} d^2 \vec{r}= \frac{2 \pi(6-9q^2)}{(1+q^2)^{\frac{7}{2}}} \nonumber \\ 
\int r^3e^{-r} e^{-i\vec{q}\cdot\vec{r}} d^2 \vec{r}= \frac{2 \pi(24+9q^2(q^2-8))}{(1+q^2)^{\frac{9}{2}}} \nonumber \\
\int r^4e^{-r} e^{-i\vec{q}\cdot\vec{r}} d^2 \vec{r}= \frac{30 \pi(8-40q^2+15q^4)}{(1+q^2)^{\frac{11}{2}}}\nonumber \\
\int r^5e^{-r} e^{-i\vec{q}\cdot\vec{r}} d^2 \vec{r}= \frac{2\pi(720-225q^2(24-18q^2+q^4))}{(1+q^2)^{\frac{13}{2}}} \nonumber \\
\int r^6e^{-r} e^{-i\vec{q}\cdot\vec{r}} d^2 \vec{r}=- \frac{630\pi(-16+7q^2(24+5q^2(q^2-6)))}{(1+q^2)^{\frac{15}{2}}}. \nonumber \\
\end{gather}

For our calculations, we use a cutoff value of $|m|,|n|\leq 30$ in Eq.~\eqref{per-int}. We omit the $\vec{q}=\vec{0}$ term in Eq.~\eqref{per-int} since it is canceled by the electron-background and background-background energies. We do not consider the self interaction energy, which is the interaction of an electron in the principal zone with its images in other zones. For a given system size, interaction and periodic boundary conditions, the  self-interaction  energy is independent of the state and thus does not affect energy differences of two states at the same filling.

\section{Certain consistency checks}
\label{sec:check}

\begin{figure}[h]
		\includegraphics[width=\linewidth]{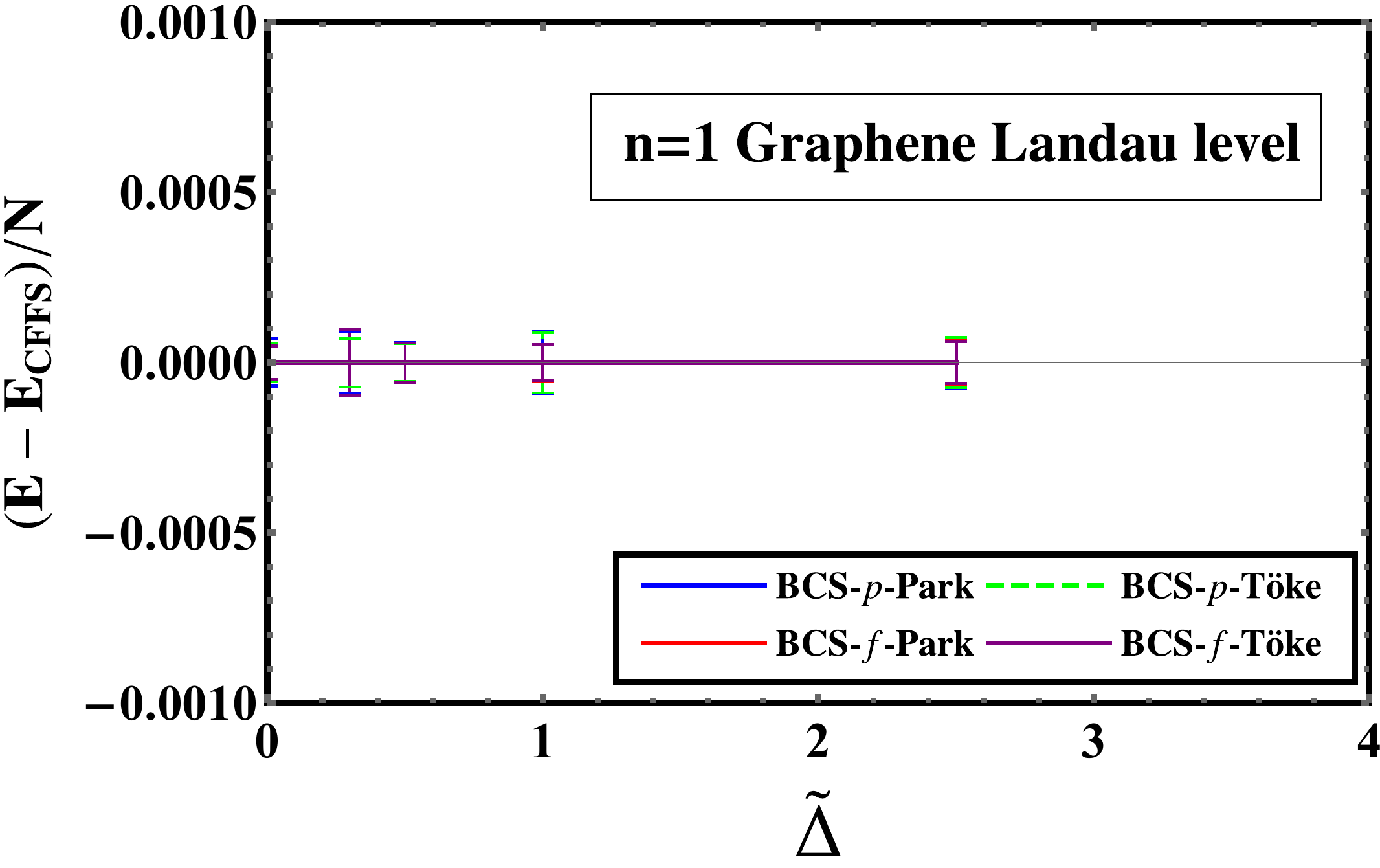}
\caption{ \label{Graphene_n_1_LL}Energy per particle for several states in $n=1$ Landau level of graphene for a system with $N=32$ particles. The energies are measured in units of $e^2/\epsilon \ell$, relative to the energy of the CFFS. The lowest energy is obtained for $k_{\rm cutoff}=k_F$ for each value of $\tilde{\Delta}$.
}
\end{figure}

\begin{figure}[h]
		\includegraphics[width=\linewidth]{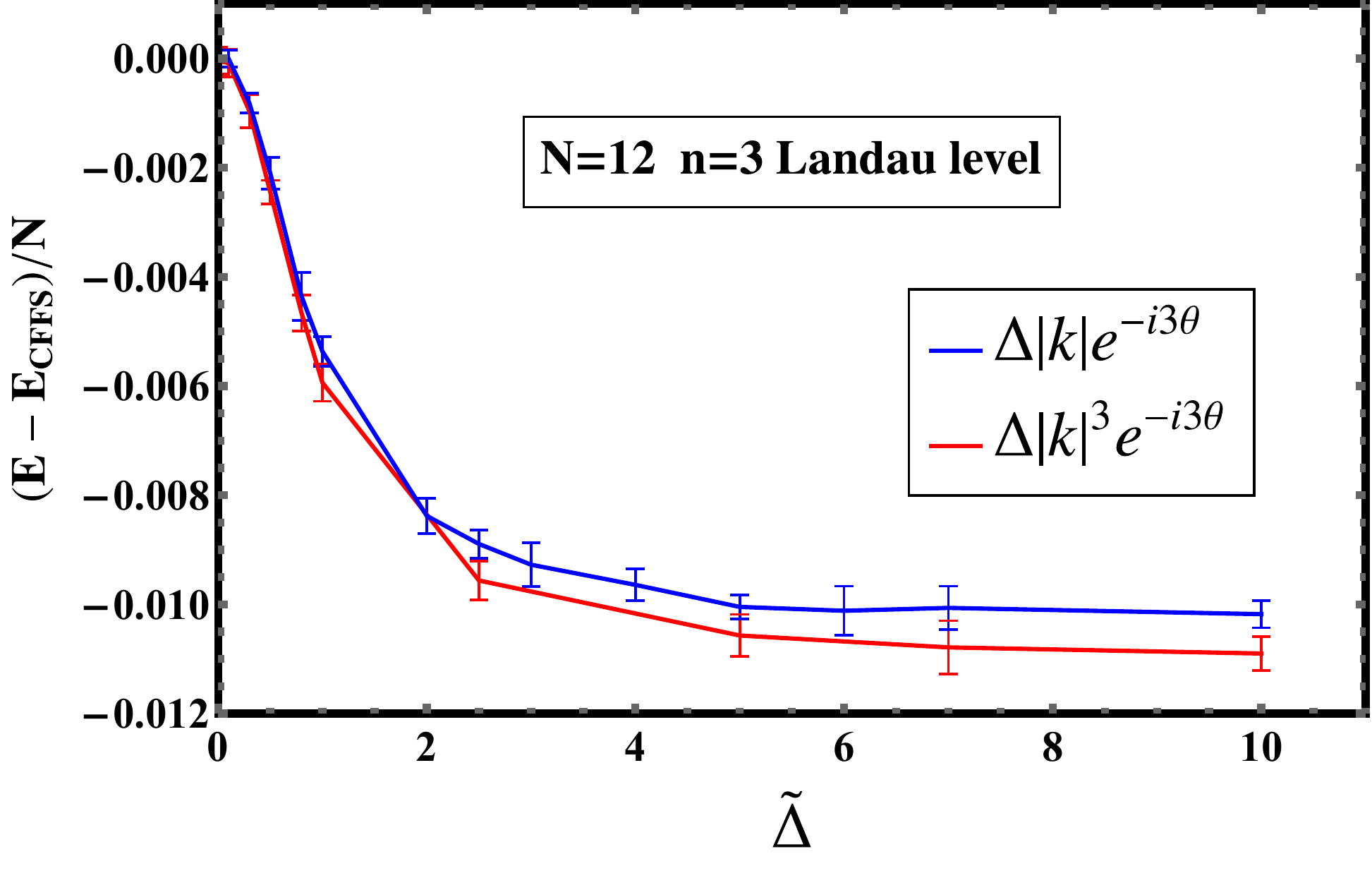}
\caption{ \label{compare}Energy per particle for the $f$-wave paired BCS states in the $n=3$ graphene Landau level with two different forms of the pairing function shown on the figure. The energies are measured relative to the energy of the CFFS.
}
\end{figure}

\begin{figure}[h]
		\includegraphics[width=\linewidth]{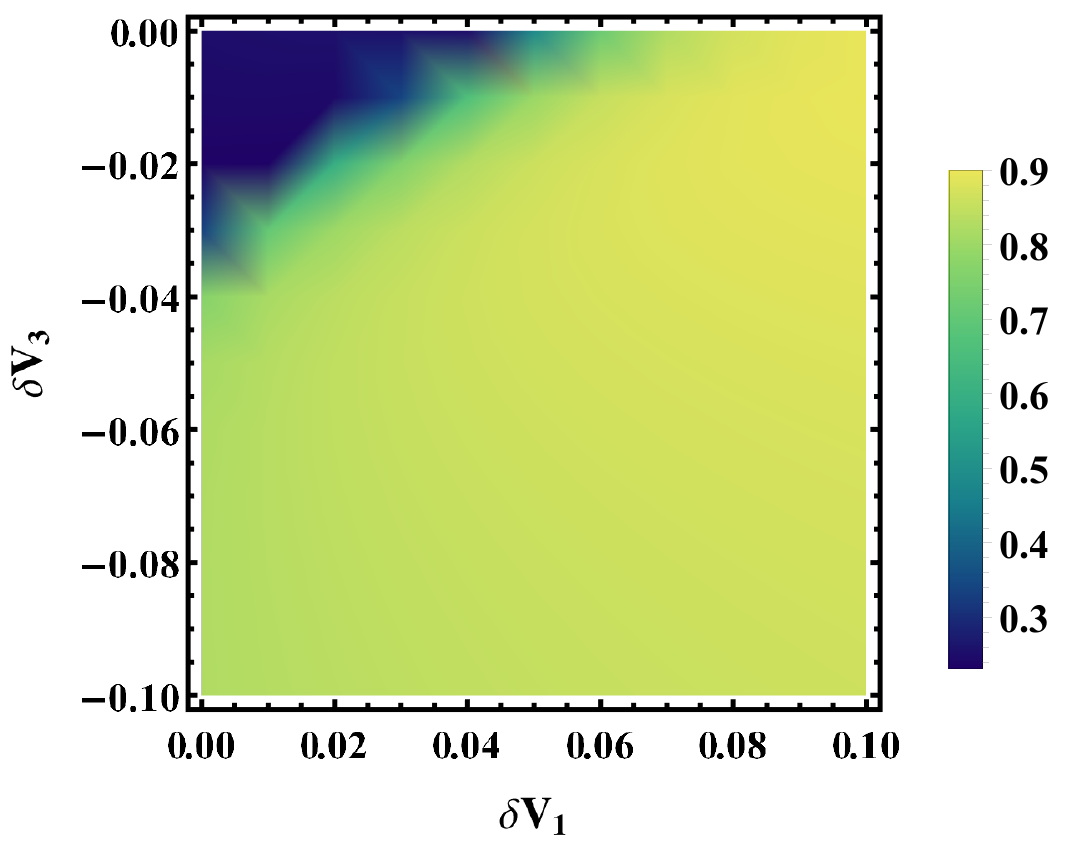}
\caption{ \label{pseudo-variation}Overlap $\left| \left<\Psi^{{\rm BCS}-f}_{\frac{1}{2}} | \Psi^{\rm{n=3}}_{\rm Ex} \right>\right|$ of the $f$-wave paired BCS states with the exact ground state in the $n=3$ Landau level as function of the variation of the first and third pseudopotentials, $\delta V_1$ and $\delta V_3$ for $N=12$. The Coulomb point is at $\delta V_1=\delta V_3=0$.
}
\end{figure}

In this section, we show certain consistency checks. We first ask how reliably our effective interactions simulate the Coulomb physics. 
As explained above, we work with effective real-space interactions (T\H oke and Park) that produce, in the lowest Landau level, the desired Haldane pseudopotentials accurately. The parameters of the effective interactions are obtained by fitting the first few odd pseudopotentials of the effective interaction in the lowest Landau level to the Coulomb pseudopotentials in the $n$th graphene LL. (This fitting procedure cannot be carried out for the Park form for the $n=2$ graphene LL since the resultant equations for the pseudopotentials do not yield a solution. Therefore, we change the pseudopotentials slightly from the Coulomb values such that for the modified pseudopotentials, the fitting procedure for the Park interaction can be carried out. This variation in pseudopotentials is not unique: we have chosen to add 0.0042 to $V_5$ of Coulomb, which is a change of 1.7\% in absolute value.)  The effective interactions are approximate, because only the first few pseudopotentials of the interaction are matched  with the Coulomb ones. However, we expect that they provide a good approximation to the Coulomb interaction, because the physics is expected to be governed by the low-$m$ pseudopotentials (which correspond to the short range part of the interaction) and because all pseudopotentials are within 2.5\% of the Coulomb values.

As a direct check of the accuracy of the Park and T\H oke interactions, we have calculated the expectation value of exact Coulomb interaction with respect to the CF-BCS wave function (at various $\tilde{\Delta}$ and $k_{\rm cutoff}$) for $N=12$ particles in the $n=2$ graphene LL. For this purpose, we first construct a Fock space representation of our CF-BCS state for a given set of parameters following the method in Ref.~\cite{Sreejith11}, which is also discussed in the Appendix of Ref.~\cite{Sharma21}. (This procedure cannot be implemented for larger $N$.) The expectation value of the Coulomb energy can now be expressed in terms of the pseudopotentials of the $n=2$ graphene LL and thus evaluated exactly. The T\H oke and exact Coulomb energies are seen in Fig.~\ref{exact-energy} to be in excellent agreement as a function $\tilde{\Delta}$ (where for each value of $\tilde{\Delta}$, we choose $k_{\rm cutoff}$ that produces the lowest energy). For the Park interaction (which is expected to be less accurate as it fits fewer pseudopotentials, and which also corresponds to a slightly modified Coulomb interaction, as mentioned in the preceding paragraph), the agreement is less good, but it still reproduces the correct qualitative behavior as a function of the variational parameters. These comparisons demonstrate that our conclusions derived from the effective interactions are valid for the Coulomb interaction.

For the $n=1$ LL of graphene, we find that the lowest energy state is obtained when $k_{\rm cutoff}=k_F$ as shown in Fig.~\ref{Graphene_n_1_LL}. This indicates in the $n=1$ LL of graphene, the lowest energy state, within the variational parameter space considered here, is the CFFS, consistent with earlier studies~\cite{Balram15c}. The overlap of the CFFS with the exact graphene $n=1$ LL state is 0.9944 for a system of 12 particles.

We also test how sensitive the conclusions are to the precise form of the gap function $\Delta_{\vec{k}}$. In the main text, we have used $\Delta_{\vec{k}}=\Delta |{\vec{k}}| e^{-i3\theta}$.  An alternative form is  $\Delta_{\vec{k}}=\Delta |{\vec{k}}|^3 e^{-i3\theta}$ \cite{Ma19}. To the extent pairing involves electrons only near the Fermi surface, we expect to recover the same state for both. We have calculated the energy per particle as a function of $\tilde{\Delta}$ for both of these forms, shown in Fig.~\ref{compare}. Both produce very similar energies, and more importantly, the choice does not alter the conclusion regarding the BCS-$f$ state being the most preferred state.

Finally, we ask how the nature of the $f$-wave paired state depends on the interaction. Fig.~\ref{pseudo-variation} shows the overlap of the CF-BCS-$f$ state and the exact ground state for $N=12$ particles in the $|n|=3$ graphene LL as the pseudopotentials $V_1$ and $V_3$ are varied by $\delta V_1$ and $\delta V_3$. The behavior is qualitatively similar to the overlap between the Jain-221 state and the exact ground state in the spherical geometry~\cite{Kim19} with the difference that for the Jain-221 state the overlap is nearly zero at the Coulomb point.

\section{Lattice Monte Carlo}
\label{sec:Monte-rev}
For our calculations, we use the lattice Monte Carlo approach introduced in Ref.~\cite{Wang19}. In this section, we present, for completeness, the central results of lattice Monte Carlo formalism necessary for our energy calculations. For a more detailed derivation, refer to Ref.~\cite{Wang19} or Appendix of Ref.~\cite{Sharma21}. Let us represent the positions of the particles as $\vec{x} \equiv (\vec{r}_1,\vec{r}_2,..,\vec{r}_N)$. In continuous Monte Carlo, the matrix elements for any operator $O(\vec{r}_i-\vec{r}_j)$ are calculated using
\begin{equation}
\frac{\bra{\Psi _1}\underset{i<j}{\sum}O(\vec{r}_i-\vec{r}_j)\ket{\Psi _2}}{\sqrt{\bra{\Psi _1}\ket{\Psi _1}\bra{\Psi _2}\ket{\Psi _2}}} = \frac{\int d^2\vec{x}\Psi_1(\vec{x})^* \Psi_2(\vec{x}) \underset{i<j}{\sum}O(\vec{r}_i-\vec{r}_j)}{\sqrt{\bra{\Psi _1}\ket{\Psi _1}\bra{\Psi _2}\ket{\Psi _2}}}.
\end{equation}
In periodic geometry, the periodic boundary conditions simplify the above calculation. The integral is now replaced by a discrete summation as follows:
\begin{equation}
\label{lat-eval}
\frac{\bra{\Psi _1} \underset{i<j}{\sum}O(\vec{r}_i-\vec{r}_j)\ket{\Psi _2}}{\sqrt{\bra{\Psi _1}\ket{\Psi _1}\bra{\Psi _2}\ket{\Psi _2}}} = \frac{\overset{\prime}{\underset{\tilde{\vec{x}}}{\sum}}\Psi_1^*(\tilde{\vec{x}}) \Psi_2(\tilde{\vec{x}}) \underset{{i<j}}{\sum}O^{\rm Lat}(\vec{r}_i-\vec{r}_j)}{\sqrt{\overset{\prime}{\underset{\tilde{\vec{x}}}{\sum}}\abs{\Psi_1}^2 \overset{\prime}{\underset{\tilde{\vec{x}}}{\sum}}\abs{\Psi_2}^2}} 
\end{equation}
where the summation on the r.h.s. is over discrete lattice points given by $\tilde{\vec{x}} \equiv (\vec{r}_1,\vec{r}_2,..,\vec{r}_N)$ with  $\vec{r}_i \in \{ (m_i\vec{L_1}+n_i\vec{L_2})/N_{\phi} | m_i,n_i \in \mathbb{Z}\}$. $\overset{\prime}{\sum}$ implies summation in the principal region of the torus. $\Psi_1$ and $\Psi_2$ are confined to the $n$th LL.

The key idea is to obtain the representation of $O^{\rm Lat}$ for a translationally invariant operator $O$. For such an operator, we can write
\begin{equation}
O(\vec{r}_i-\vec{r}_j) = \frac{1}{L^2\rm{Im} (\tau)}\sum_{\vec{q}} \tilde{O}(\vec{q})e^{i\vec{q}\cdot (\vec{r}_i-\vec{r}_j)}.
\end{equation}
with $\tilde{O}(\vec{q})$ being the Fourier transform of $O(\vec{r})$. The summation is over all discrete $\vec{q}$ allowed by periodic boundary conditions:
\beq
\label{q-vec}
\vec{q}_{m,n} = \left( {2\pi m \over L}, -{2\pi \tau _1 m \over L \tau _2} +{2\pi  n \over L \tau _2} \right)
\eeq
Splitting the coordinates and wave functions into Landau orbits and the guiding center part, we can write:
\begin{widetext}
\beq
\label{guiding-center}
\bra{\Psi _1}O(\vec{r}_i-\vec{r}_j)\ket{\Psi _2} = \frac{1}{L^2\rm{Im} (\tau)}\sum_{\vec{q}} \tilde{O}(\vec{q})f_n^2(\vec{q})\bra{\Psi _1^{GC}}e^{i\vec{q}\cdot (\vec{R}_i -\vec{R}_j)} \ket{\Psi _2^{GC}}
\eeq
\end{widetext}
where $\Psi_1^{GC}$ and $\Psi_2^{GC}$ are the guiding center parts of the wave functions in guiding center coordinates $\vec{R}_i$, and  $f_n(\vec{q})=e^{-\frac{\abs{\vec{q}}^2}{4}}L_n \left (\frac{\abs{\vec{q}}^2}{2}\right )$ is the form factor in the $n$th LL ($L_n(q)$ is the Laguerre polynomial).  The summation over $\vec{q}$ in Eq.~\eqref{guiding-center} can be divided into two parts: the first BZ and the rest of the momentum space. Since the later part can be incorporated into the summation of the first BZ using the periodic boundary conditions, we can thus rewrite Eq.~\eqref{guiding-center} as:
\begin{multline}
\bra{\Psi _1}O(\vec{r}_i-\vec{r}_j)\ket{\Psi _2}\\ = \frac{1}{L^2\rm{Im} (\tau)}\sum_{\vec{q}}^{\prime} O^{GC}(\vec{q})\bra{\Psi _1^{GC}}e^{i\vec{q}\cdot (\vec{R}_i -\vec{R}_j)} \ket{\Psi _2^{GC}}
\end{multline} 
where $\displaystyle \sum^{\prime}_{\vec{q}}$ represents summation in the first BZ. $O^{GC}(\vec{q})= \sum_{\vec{q}'}\tilde{O}(\vec{q}+N_{\phi}\vec{q}')f_n^2(\vec{q}+N_{\phi}\vec{q}')$ is defined only in the first BZ but it incorporates the short range part using the $\vec{q}'$ summation. The allowed values of $\vec{q}'$ are also given by Eq.~\eqref{q-vec}. In our numerical calculations, we consider a cutoff on the number of $\vec{q}'$ in the summation. We choose $\abs{m},\abs{n} \leq 50$, which guarantees convergence.
Using the properties of Fourier transform on a lattice and boundary conditions, we obtain the expression for a two body operator $O^{\rm Lat}(\vec{r}_i-\vec{r}_j)$ \cite{Wang19,Sharma21}:
\beq
O^{\rm Lat}(\vec{r}_i-\vec{r}_j) = \frac{1}{2\pi N_{\phi}} \sum_{\vec{q}}^{\prime} \Bigg ( \frac{[f_n(0)]_{N_{\phi}}}{[f_n(\vec{q})]_{N_{\phi}}} \Bigg ) ^2 O^{GC}(\vec{q})e^{i\vec{q}.(\vec{r}_i-\vec{r}_j)} \nonumber
\eeq
where $[f_n(\vec{q}_{l,m})]_{N_{\phi}} = \sum_{\vec{q'}_{j,k}}f_n(\vec{q}_{l,m}+N_{\phi}\vec{q'}_{j,k})e^{i(-k\phi_1+j\phi_2)} \cross (-1)^{lk-mj+N_{\phi}jk}$ is the compactified form factor. We can also implement the lattice Monte Carlo to calculate the overlap between two states in Eq.~\eqref{lat-eval}, which is given by
\beq
\frac{\bra{\Psi _1} \ket{\Psi _2}}{\sqrt{\bra{\Psi _1}\ket{\Psi _1}\bra{\Psi _2}\ket{\Psi _2}}} = \frac{\overset{\prime}{\underset{\tilde{\vec{x}}}{\sum}}\Psi_1^*(\tilde{\vec{x}}) \Psi_2(\tilde{\vec{x}}) }{\sqrt{\overset{\prime}{\underset{\tilde{\vec{x}}}{\sum}}\abs{\Psi_1}^2 \overset{\prime}{\underset{\tilde{\vec{x}}}{\sum}}\abs{\Psi_2}^2}} 
\eeq
The Metropolis algorithm is set up in the same way as that of the continuous case. For our calculations, we use 50,000 iterations for thermalization, and 10 samples with 10,000,000 iterations for Monte Carlo averaging.

\bibliography{biblio_fqhe.bib}
\end{document}